\documentclass[review,5p]{elsarticle}

\usepackage{lineno}
\modulolinenumbers[5]
\usepackage[varg]{txfonts}
\usepackage{graphicx,subfigure}
\usepackage{epstopdf}
\usepackage{txfonts}
\usepackage{natbib}
\usepackage{multirow}
\usepackage{threeparttable}
\usepackage{booktabs}
\usepackage{colortbl}
\definecolor{mygray}{gray}{.9}
\usepackage{ulem} 
\usepackage{color}
\usepackage[colorlinks,linkcolor=blue,anchorcolor=green,citecolor=red]{hyperref}
\usepackage{array}
\renewcommand\arraystretch{0.7}
\usepackage{longtable}

\biboptions{authoryear}

\begin{document}

\begin{frontmatter}

\title{Efficient Fermi Source Identification with Machine Learning Methods}
\author[phypd,astrogz,infnpd,gzhu]{H. B. Xiao}
\author[engpd,infnpd,gzhu]{H. T. Cao\corref{corrauth1}}
\author[astrogz,gzhu]{J. H. Fan\corref{corrauth2}}
\author[astrogz,astrochina,stapd]{D. Costantin}
\author[gzhu]{G. Y. Luo}
\author[phypd,astrogz,infnpd,gzhu]{Z. Y. Pei}

\cortext[corrauth1]{Corresponding author: haitao.cao@phd.unipd.it}
\cortext[corrauth2]{Corresponding author: fjh@gzhu.edu.cn}

\address[phypd]{Department of Physics and Astronomy "G. Galilei", University of Padova, Via Marzolo, 8, Padova PD, 35131, Italy.}
\address[astrogz]{Center for Astrophysics, Guangzhou University, No. 230, Wai Huan Xi Road, Guangzhou Higher Education Mega Center, Guangzhou, 510006, China.}
\address[engpd]{Department of Information Engineering, University of Padova, Via Gradenigo, 6/B, Padova PD, 35131, Italy.}
\address[infnpd]{Istituto Nazionale di Fisica Nucleare, Via Marzolo, 8, Padova PD, 35131, Italy.}
\address[gzhu]{School of Physics and Electronic Engineering, Guangzhou University, No. 230, Wai Huan Xi Road, Guangzhou Higher Education Mega Center, Guangzhou, 510006, China.}
\address[astrochina]{Astronomy Science and Technology Research Laboratory of Department of Education of Guangdong Province, Guangzhou, 510006, China.}
\address[stapd]{Department of Statistical Science, University of Padova, Via Battisti, 241, Padova PD, 35121, Italy.}

\begin{abstract}
{In this work, Machine Learning (ML) methods are used to efficiently identify the unassociated sources and the Blazar Candidate of Uncertain types (BCUs) in the \textit{Fermi}-LAT Third Source Catalog (3FGL).} 
{The aims are twofold: 1) to distinguish the Active Galactic Nuclei (AGNs) from others (non-AGNs) in the unassociated sources; 2) to identify BCUs into BL Lacertae objects (BL Lacs) or Flat Spectrum Radio Quasars (FSRQs).}
{Two dimensional reduction methods are presented to decrease computational complexity, where Random Forest (RF), Multilayer Perceptron (MLP) and Generative Adversarial Nets (GAN) are trained as individual models. In order to achieve better performance, the ensemble technique is further explored.}
{It is also demonstrated that grid search method is of help to choose the hyper-parameters of models and decide the final predictor, by which we have identified 748 AGNs out of 1010 unassociated sources, with an accuracy of 97.04\%. Within the 573 BCUs, 326 have been identified as BL Lacs and 247 as FSRQs, with an accuracy of 92.13\%.}

\end{abstract}


\end{frontmatter}


\section{Introduction}
Active Galactic Nuclei (AGNs) have been a hot topic in astrophysics since their discovery in 1960s, while their nature still remains uncertain. The study of AGNs is always one of the paramount issues.  For instance, it provides important clues for the evolution and structure of galaxies \citep{Hopkins2006} and the whole Universe in the reionization era. AGNs have also been used to constrain the cosmological constant and the baryon density $\Omega_{\rm B}$ \citep{Varshalovich2012}.

In some of AGNs, they contain an energetic jet, which emits from the center of a Supermassive Black Hole (SMBH) and is surrounded by an accretion disk \citep{Urry1995}. The jets emissions are dominated by nonthermal radiation, with spectrum spanning from radio to $\gamma$-ray band. The Spectral Energy Distribution (SED) of jet emissions shows a two-hump structure. The lower-energy hump, peaked between millimeter band and soft X-ray band, is attributed to the synchrotron emission produced by electrons in the jet. And, the higher-energy hump, peaked in the MeV-GeV range, is mainly due to inverse Compton scattering according to leptonic model. While in a hadronic senario, protons are assumed to be accelerated along with electrons, and contribute much of the high-energy component via photopion interactions \citep{Dimitrakoudis2012,Bottcher2009}.

AGNs standard model also shows that different types of AGNs result from different viewing angles between the line of sight and the jet direction. In particular, the sources can be classified as blazars if their jets are pointing towards the observer. Blazar is an extreme subclass of AGNs, since it shows high and fast flux variability, coupled with rapidly varying polarization. It can be divided into two subclasses: BL Lacertae objects (BL Lacs) and Flat Spectrum Radio Quasars (FSRQs). The BL Lac shows weak or no emission lines, while the FSRQ shows strong emission lines.

\textit{Fermi}-LAT (Large Area Telescope on-board Gamma-Ray Space Telescope, see \citealt{Atwood2009}) is a powerful imaging high-energy $\gamma$-ray telescope launched in 2008 and spans over the energy ranges from about 20 MeV to more than 300 GeV. Its field of view covers about 20\% of the sky at any time and the whole sky in every three hours through a continuous scan. With respect to the previous telescope (e.g., EGRET, the Energetic Gamma-Ray Experiment Telescope, see \citealt{Fichtel1993}), \textit{Fermi}-LAT shows better energy resolution, better angular resolution and wider effective area in both low-energy and high-energy bands\footnote{Check \url{http://www.slac.stanford.edu/exp/glast/groups/canda/lat\_Performance.htm} for more performance of \textit{Fermi}-LAT.}. And \textit{Fermi}-LAT collaboration provides point source catalogues, which have been used to study the properties of AGNs and their subclasses.

Our work focused on utilizing the third \textit{Fermi}-LAT source catalog (3FGL), which covers the four-year observation data of the \textit{Fermi}-LAT. 3FGL contains 3034 sources, among which 1744 belong to AGN class. Within the AGN sources, there are 573 Blazars that haven't been tagged as BL Lacs or FSRQ. These uncertain Blazars are also named as Blazar Candidate of Uncertain types (BCUs). Besides, the 3FGL remains 1010 unassociated sources \citep{3FGL}. These BCUs and unassociated sources result in an incomplete catalog, which decreases the significance of observations. However, a certain source sample can be of great help to study its high energy properties, e.g., variability and radiative process. Therefore, efficient algorithms are desirable to handle the incompleteness problem of 3FGL.

Many efforts have been dedicated to this issue \citep{Hassan2013,Doert2014,Cavuoti2014,Kang2019}. For instance, \citet{Parkinson2016} used two Machine Learning (ML) models, i.e., Random Forest (RF) and Logistic Regression (LR) to identify 1008 unassociated 3FGL sources, of which 893 sources were in agreement with both methods. But the authors probably ignored the precision of predicted AGNs, which will be one of the contributions in our work.

Since variability is one of the characterizing properties of Blazar \citep{Paggi2011}, \citet{Chiaro2016} utilized Blazar Flaring Patterns (B-FlaP) as the inputs of an Artificial Neural Networks (ANN) to identify the 573 BCUs in 3FGL, while they have 77 still remained uncertain. Similarly, \citet{Lefaucheur2017} also left 45 BCUs unprocessed. We will further deal with these remained BCUs and discussed via comparisons in this paper.

However, the increasing number of sources, coming both from current and next-generation telescopes, requires more powerful classifiers. In this work, different ML methods are explored to design an efficient algorithm for further identifying 3FGL unassociated sources and BCUs. The improvements in identification efficiency mainly focus on two aspects:
\begin{enumerate}[1)]
\item We introduce two dimensionality reduction methods acting as the data preprocessing step to remove the redundancy and noise of data and decrease computational complexity.
\item The independently trained models are aggregated to be the ensemble model for strong generalization capability and better performance on unseen samples.
\end{enumerate}
Furthermore, grid search method is used to record the hyper-parameters achieving the best performance and to choose the final predictors acting on the identification tasks. Finally, the evaluation of models on multi-splitting datasets demonstrates the effectiveness of proposed method on the identification of the unassociated sources and BCUs in 3FGL.

The paper is organized as follows. Section \ref{sect_ml} presents a brief introduction of ML methods we used. Section \ref{sect_proposed} describes our proposed method in detail. The experiments and results are reported in Section \ref{sect_experiment}. Further discussions and conclusions are presented in Section \ref{sect_discussion} and \ref{sect_conclusion}, respectively.

\section{Machine Learning Methods}
\label{sect_ml}
As ML increases success in many computer science and engineering applications, it is gaining popularity in the astronomical subjects as well, for instance to complete the 3FGL data catalog \citep{Parkinson2016,Chiaro2016}. Similar to the image classification in computer science, the source identification task could be conducted by building a classifier.

\subsection{Dimensionality Reduction}
High dimensionality is one of the best-known problems when processing and analyzing big data, due to the high costs of the hardware resources and computing time. Researches have shown that the experimental data are highly redundant, leaving room for improvement through the design of efficient algorithms and the reduction of the dimensionality \citep{Bakshi1993,Bengio2009,Glorot2011}. Lower dimensionality will also lead to a decreased computational complexity, which has become a preferred avenue nowadays \citep{He2018,Sellami2018,Kuang2018}.

One of the widely used techniques for reducing dimensionality is Feature Selection (FS). With this method, we select a subset of features that are most relevant to the task. It improves computational efficiency and reduces the generalization error of model by removing irrelevant features or noise \citep{Sebastian2015}. Feature Importance (FI) is commonly used to measure the feature significance. It is usually calculated from the average impurity decrease of the criterion brought by the particular feature in RF models. We pick out the features with high FI values, while abandon others whose FI values are low. RF was firstly proposed by \citet{Kam1995} and since then it has become a powerful ML method for classification and prediction tasks despite its simple mechanism \citep{Breiman2001,Parkinson2016}. A RF consists of many Decision Trees (DTs) and manages to reduce both the bias and variance of the tree models, which is the idea of Ensemble Learning (EL) \citep{Dietterich2002}. 

Another strategy for dimensionality reduction is Feature Extraction (FE). This method transforms or projects the original data into a new feature space, in which the data will be represented by other types of variables. Principal Component Analysis (PCA) is an unsupervised linear transformation technique for dimensionality reduction. With PCA, the data is transformed from its original coordinate system to a new coordinate system. The first new axis is chosen in the direction of the largest variance in the data. Then the second axis is orthogonal to the first axis and in the direction of the second largest variance. This procedure is repeated until the number of features is met. The majority of the variance is contained along the first few axes. Therefore, the rest of the axes can be neglected and as a result the dimensionality of data can be reduced \citep{Peter2012}.

\subsection{Neural Network Classifiers}
After data preprocessing, classifiers are built for data analysis. Neural Network (NN) is an important aspect of ML methods. It has been successfully applied in different classification tasks within the astrophysical content \citep{Chiaro2016,Salvetti2017}. It usually consists of one input layer, several hidden layers, and one output layer. One-time NN training process includes Forward Propagation (FP) and Back Propagation (BP), but a good NN model always requires training many times. In FP the data flows from input to output layer, while in BP the flow has a converse direction.

FP aims to calculate the model prediction errors based on a predefined loss function, such as the classical cross-entropy function. A lower error always means a higher accuracy score for the classification task, and indicates a better classifier. In addition to the cross-entropy function, in our analysis we also considered some other non-classical loss functions listed in \citet{Janocha2016}.

BP algorithm was proposed by \citet{Rumelhart1988} to update the parameters of NN model. Starting from the predicted error obtained in output layer, the errors of each hidden layer are successively calculated. With these errors, the model parameters are then updated by different optimization algorithms \citep{Duchi2011,Sutskever2013}.

The layers of a Multiplayer Perceptron (MLP) model are often fully connected to each other, which are also known as Fully-connected Layers (FcLs) \citep{Cavuoti2014}. Another type of NN model, Generative Adversarial Nets (GAN) that firstly proposed by \citet{Goodfellow2014}, can also be constructed by FcLs. It includes a discriminator and a generator, between which an adversarial game takes place: the discriminator is trained to tell the samples coming from the generator, while the generator is trained to produce samples that cannot be recognized by the discriminator.

\citet{Odena2017} and \citet{Salimans2016} successfully built a semi-supervised classifier with GAN model. They labeled the samples produced from the generator as one class, namely the fake class, while the samples that came from the dataset were labeled as true. As the generator was trained to approximate true samples, the discriminator was trained to identify fake samples. Moreover, since the true samples had distinct known classes as well, the discriminator was additionally trained to predict these known classes. Their work obtained good results on some public image datasets.

\subsection{Ensemble Methods}
Traditionally, dataset is divided into two sub-datasets: training dataset for model training, and test dataset for model evaluation. The so-called over-fitting, referred to the phenomenon that a model performs well on training dataset while badly on test dataset, is a common issue in ML field. It actually indicates the weak generalization capability of a model. That is, a model with strong generation capability will achieve satisfactory performance on unseen samples, as expected from classification tasks. The Ensemble Method (EM), proposed by \citet{Freund1995}, aggregates independently trained ML models into an ensemble predictor and reduces both the bias and the variance of individual models, which is of help to alleviate the over-fitting problem.

Two strategies are usually used to aggregate models: hard and soft. The former counts the prediction of each individual model and gives the class that gets the most votes \citep{Zhang2014}. The drawback is that it cannot estimate the likelihood of predicted class. In other words, it cannot tell how certain the class of the predicted result is. Therefore, this strategy will not be considered in our work.

While the latter identifies samples based on the weighted likelihood of predictions coming from the individual models \citep{Jimenez1998}. These weights indicate the contributions of models and they are decided by the performance of individual models, i.e., the model with higher accuracy score will hold heavier weight. Obviously, the soft ensemble strategy requires every individual model to have the ability of estimating the likelihood of prediction. This aggregation strategy will be used in our analysis.

\section{Proposed Method}
\label{sect_proposed}
\begin{figure}[htbp]
\centering
\subfigure[]{
\includegraphics[scale=0.4]{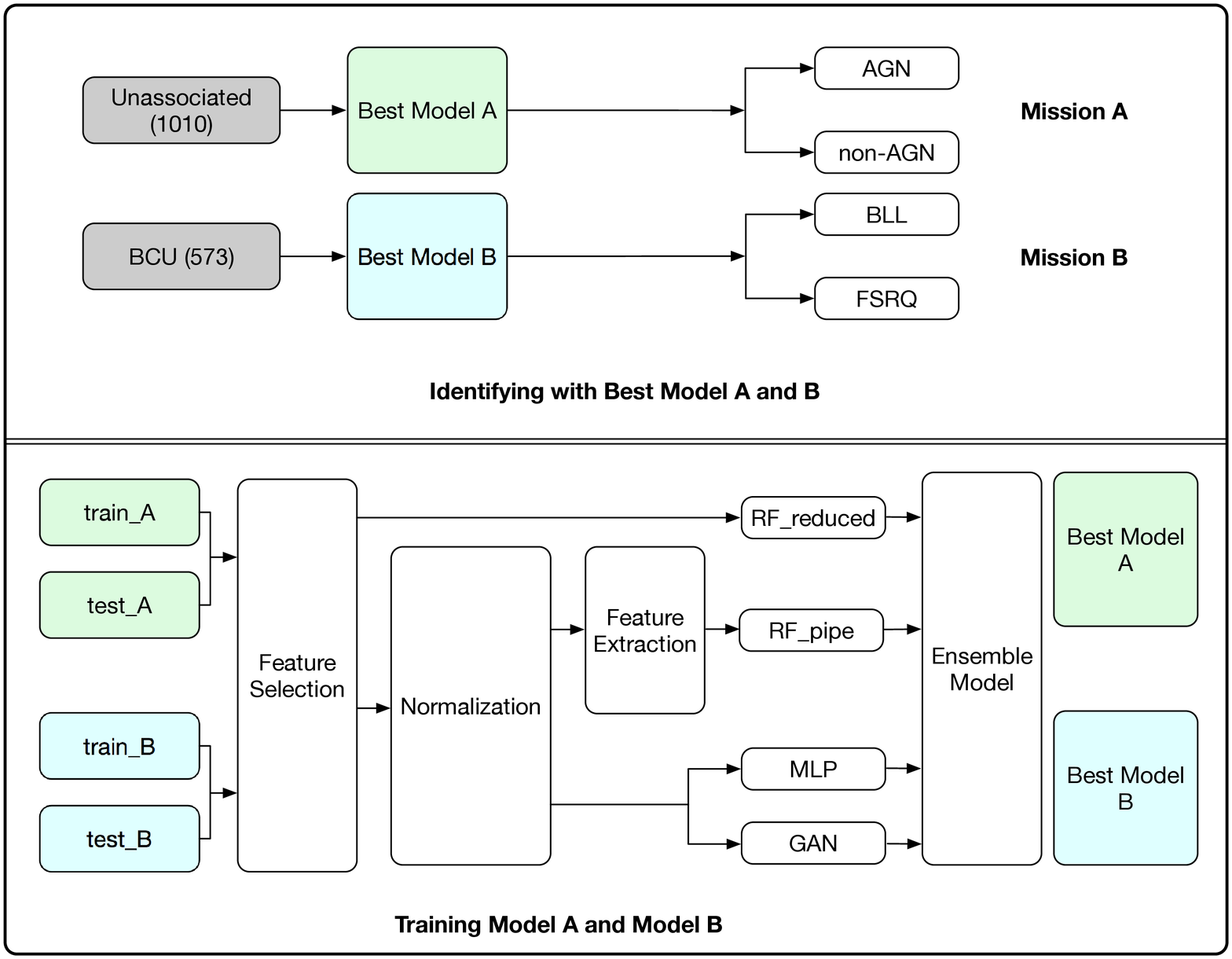}
\label{fig_model}
}
\subfigure[]{
\includegraphics[scale=0.4]{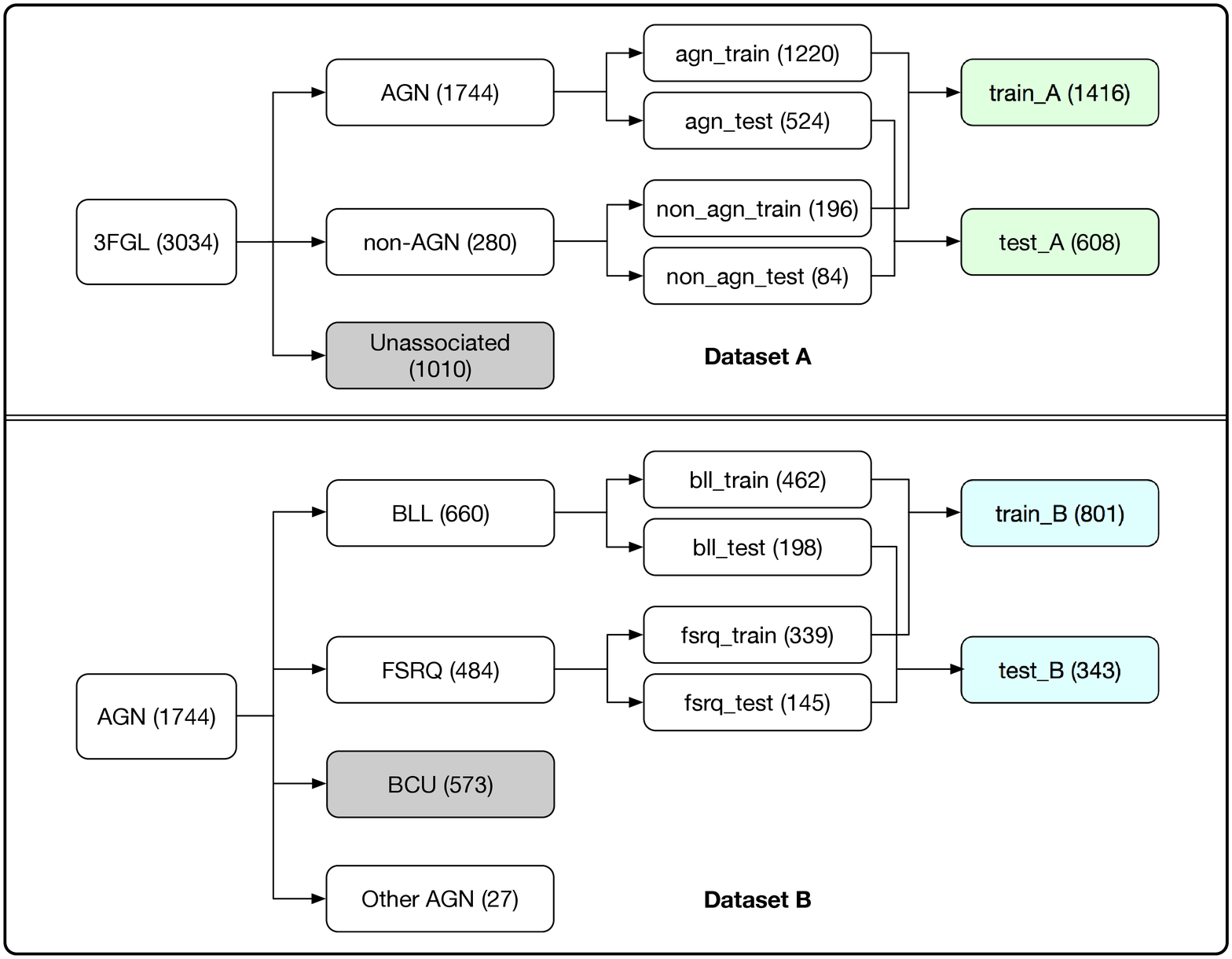}
\label{fig_dataset}
}
\caption{(a) Illustration of missions (top) and models (bottom). (b) Datasets of 3FGL sources split for Mission A (top) and B (bottom).}
\end{figure}

There are two efforts in our work: Mission A and Mission B, shown in the top panel of Figure \ref{fig_model}. We aim to build a Best Model A able to identify AGNs versus non-AGNs for Mission A, and a Best Model B able to distinguish FSRQs and BL Lacs for Mission B. The Best Model in each mission is chosen from five candidate models: `RF\_reduced', `RF\_pipe', `MLP', `GAN' and `Ensemble Model', shown in the bottom panel of Figure \ref{fig_model}. The candidate model achieving best performance on test dataset will be taken as the Best Model used to accomplish the identification task.

Besides, only one-time dataset split may result in a biased model, thus we will randomly split the dataset into training dataset and test dataset several times and repeat the procedure, which is the so-called hand-out evaluation method. In this way, the model will be evaluated more than once and it will achieve strong generalization capability if its performance on each splitting test dataset is comparable.

\subsection{Data Preparation}
If we assume that ML models, in particular, the NN models, have the capability of finding underlying function relationship among different features, then it is not needed to use some predefined parameters such as hardness ratio in \citet{Parkinson2016} and flux ratio in \citet{Mirabal2016}. We will not consider these parameters in our work. Instead, we directly use FS method to select the subset of features that are most relevant to our tasks. Experimental Section will give more details on the feature characteristics.

\subsection{Method Scheme}
Training the ML models is a challenge. One of the main difficulties comes from the choice of so many hype-parameters, such as the depth, criterion and the number of trees in RF model and the number of hidden layers in MLP and GAN models. There is not a general rule to choose the suitable set of hyper-parameters. In our work, grid search will be used in every step. The procedures for two missions are the same. We will consider different candidate models coming from RF, MLP, GAN and ensemble models.

Following the scheme of Figure \ref{fig_model}, the first step of the procedure is FS. Based on different levels of FI values, several subsets of features are generated. If the FI values are higher than a set level then these features are selected into a subset. The data within the features of a subset will be fed to next step.

The scheme is then divided into two possibilities: normalization or not. Normalization ensures that values of different features range in the same scale. We will use four normalization methods, namely `Standardize', `Normalize', `Minmax' and `MaxAbs' \citep{Pedregosa2011}. Therefore, RF has two different processes for input features: the one not normalized, i.e., `RF\_reduced', or the normalized one followed by FE step, i.e., `RF\_pipe'.

Whereas MLP and GAN models are fed just with the normalized features. The reasons why we don't consider FE step here come from two sides. On one hand, PCA is a linear dimensionality reduction technique. It is not appropriate for complex problems with nonlinear correlation structures \citep{Zhang2000}. While NN models have the nonlinear fitting capability for dealing with these problems. If the FE step is added ahead of NN models, the nonlinear correlation structures of data may be destroyed and cannot be seen by NN models. As a result, the identification performance will degrade. On the other hand, when performing dimensionality reduction with PCA, the discriminative information that distinguishes one class from another one may exist in the low variance components and may be removed improperly, which makes performance worse.

The last candidate model comes from the soft ensemble model, which is aggregated by the four weighted individual models. The sum of weights is 1. All of the possible combinations of weights in two decimal places will be tested and the one which gives the best performance will be recorded.

\subsection{Evaluation of Performance}
For identification tasks, there exist three metrics for performance evaluation: Sensitivity, Precision, and Accuracy. They can be defined by the form of confusion matrix, which consists of four prediction cases:
\begin{itemize}
\item True Positive (TP): the number of positive samples correctly predicted to be positive;
\item False Positive (FP): the number of negative samples wrongly predicted to be positive;
\item False Negative (FN): the number of positive samples wrongly predicted to be negative;
\item True Negative (TN): the number of negative samples correctly predicted to be negative.
\end{itemize}
Then the aforementioned three metrics can be written as:
$$\mathrm{Sensitivity}=\frac{\mathrm{TP}}{\mathrm{TP+FN}}$$
$$\mathrm{Precision}=\frac{\mathrm{TP}}{\mathrm{TP+FP}}$$
$$\mathrm{Accuracy}=\frac{\mathrm{TP}+\mathrm{TN}}{\mathrm{TP}+\mathrm{FP}+\mathrm{TN}+\mathrm{FN}}$$
In our experiments, the performance of models in Mission B will be evaluated by accuracy. While in Mission A, both the accuracy and precision will be considered.

\section{Experiments and Results}
\label{sect_experiment}
The experiments were set up on the Ubuntu 18.04 system. To speed up the computing, RF programs were distributed on multi-CPUs with scikit-learn (version 0.20) \citep{Pedregosa2011}. While NN models were accelerated by GPU with TensorFlow (version 1.10) \citep{Abadi2016}. In the experiments, the CPU (Central Processing Unit) was Intel Core i5-8400 and the GPU (Graphic Processing Unit) was NVIDIA GTX1050Ti with 768 CUDA cores.

\subsection{Source Samples}
In this work, we used the 3FGL data downloaded from Space Science Data Center (SSDC)\footnote{\url{http://www.ssdc.asi.it/fermi3fgl/}}. To proceed with ML methods, we randomly split all the known entries into two datasets: training dataset (70\% of the entries) and test dataset (the remaining 30\%). As already mentioned in Section \ref{sect_proposed}, to test the generalization ability of models, we repeatedly split the dataset five times. As the top panel of Figure \ref{fig_dataset} shows, Dataset A had 2024 associated sources and it was divided into the two subclasses: 1744 AGNs and 280 non-AGNs. 70\% of them, i.e., 1220 AGNs and 196 non-AGNs, formed the training dataset. While the remaining 30\%, with 608 sources (524 AGNs and 84 non-AGNs) was used to evaluate the performance of trained models. The same methodology was applied to Dataset B, as shown in the bottom panel of Figure \ref{fig_dataset}. In the experiments, 21 features were picked out from the 32 attributes of the SSDC 3FGL data for the ML methods, as shown in Table \ref{tab_features}.

\subsection{Data Analysis}
In FS step, the FI values of 21 features were calculated with a RF model whose hyper-parameters can be found in Table \ref{tab_fs_parameters}. With two optional number of DTs and two criterions, we got four FI results.

\begin{figure}[h]
\centering
\subfigure[]{
\includegraphics[scale=0.4]{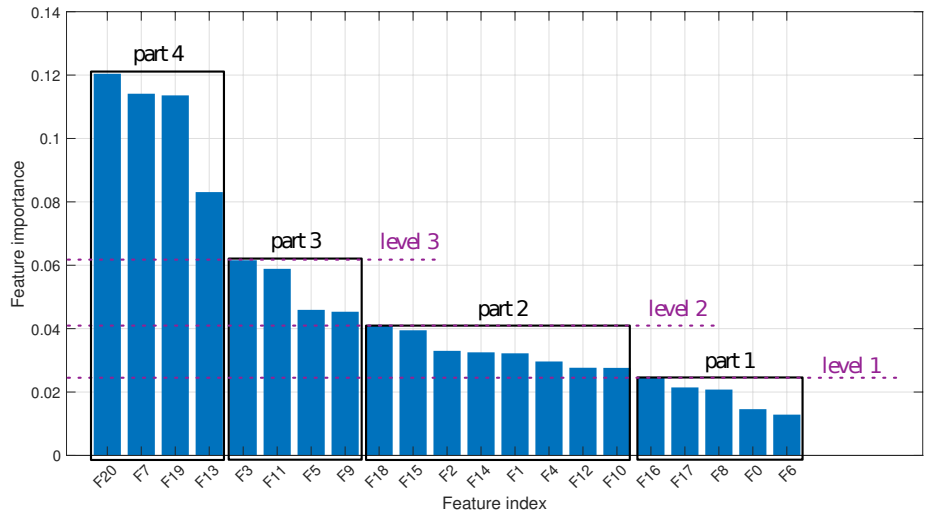}
\label{fig_fs_exam_a}
}
\subfigure[]{
\includegraphics[scale=0.4]{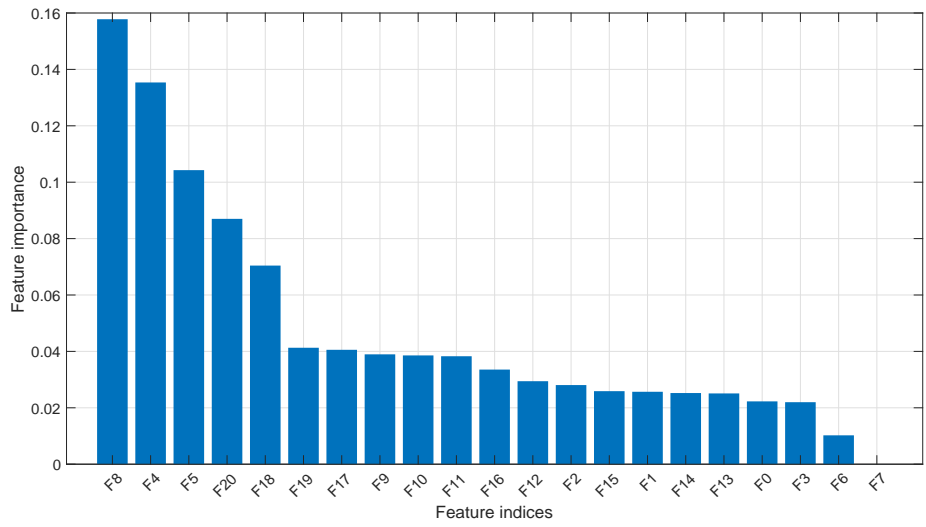}
\label{fig_fs_exam_b}
}
\caption{The descending FI values in the (a) third split case of Dataset A, and (b) second split case of Dataset B.}
\label{fig_fs_exam}
\end{figure}

Figure \ref{fig_fs_exam_a} displays the FI values in the third split case of Dataset A when we chose 30000 DTs and set `entropy' as the criterion. Empirically, the features were divided into four groups roughly based on three importance levels. The first few groups that contain most unimportant features could be removed and then feature-removed datasets were prepared for next step. In Figure \ref{fig_fs_exam_a}, there were three groups of features could be removed, i.e., (part 1), (part 1, part 2) and (part 1, part 2, part 3). Moreover, the case of the whole 21 features without any removing was also tested in our work for comparison. All the removable features in each split case are listed in the second and third columns of Table \ref{tab_fs_fe}.

Due to empiricism, the feature division can be different and some features may be ambiguously divided. For instance, feature `F1' would move from part 1 to part 2 when we changed the criterion from `gini' to `entropy' in the case of 10000 DTs. But we found it didn't affect the predicted results a lot, thus we ignored the differences of different feature divisions.

Similarly, in FE step, the components projected from the feature-removed dataset were also divided into several parts according to different variance levels. Only the parts characterized with high variance values were kept. In the experiments we found that, for a particular feature-removed dataset, the number of principal components was the same in spite of different normalization methods ahead of FE step. In other words, only the FS method influenced the number of principal components. However, we didn't remove the normalization step since it differed the results of `RF\_pipe' models.

\begin{figure}[htbp]
\centering
\subfigure[]{
\includegraphics[scale=0.4]{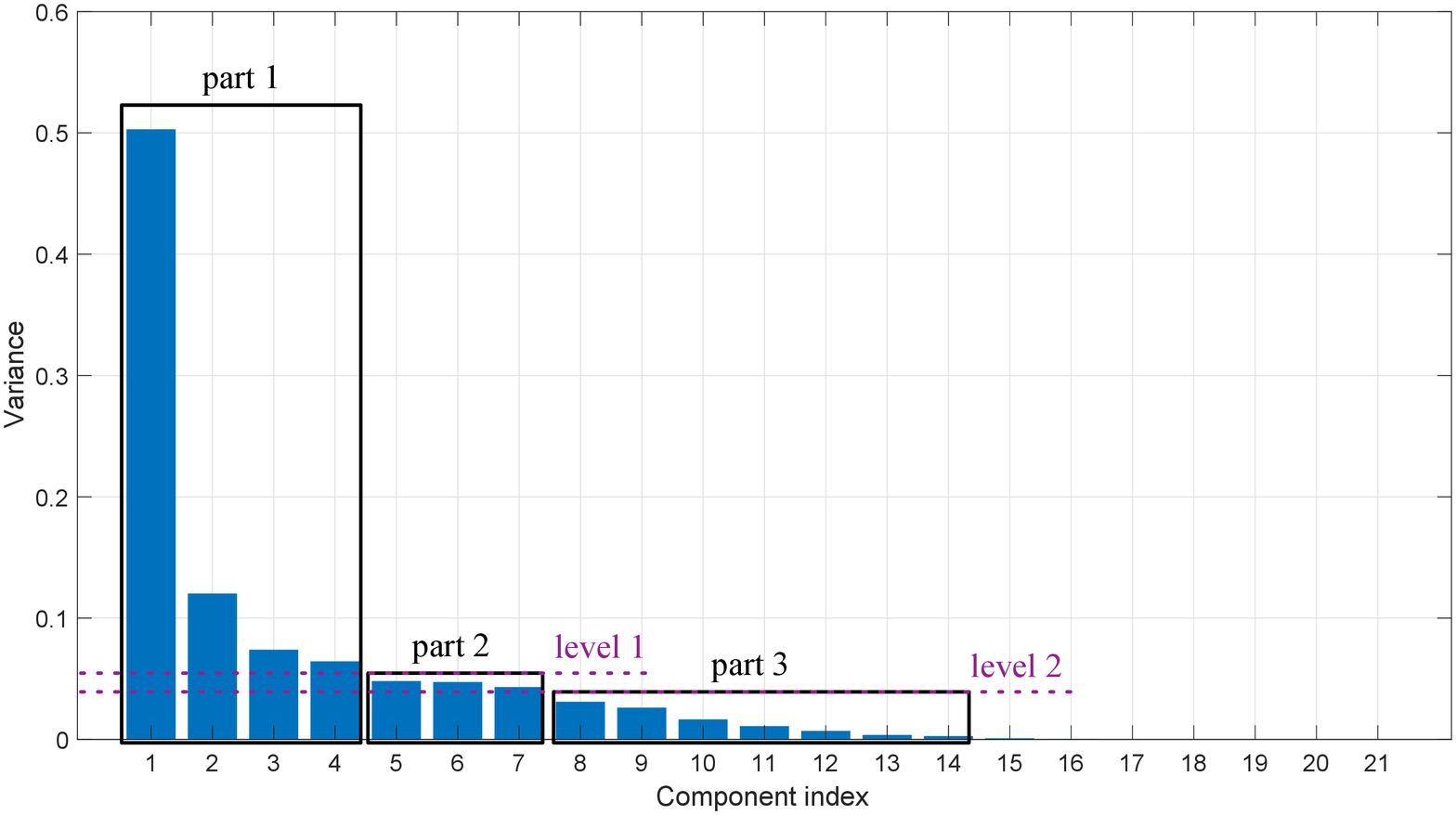}
\label{fig_fe_exam_a}
}
\subfigure[]{
\includegraphics[scale=0.4]{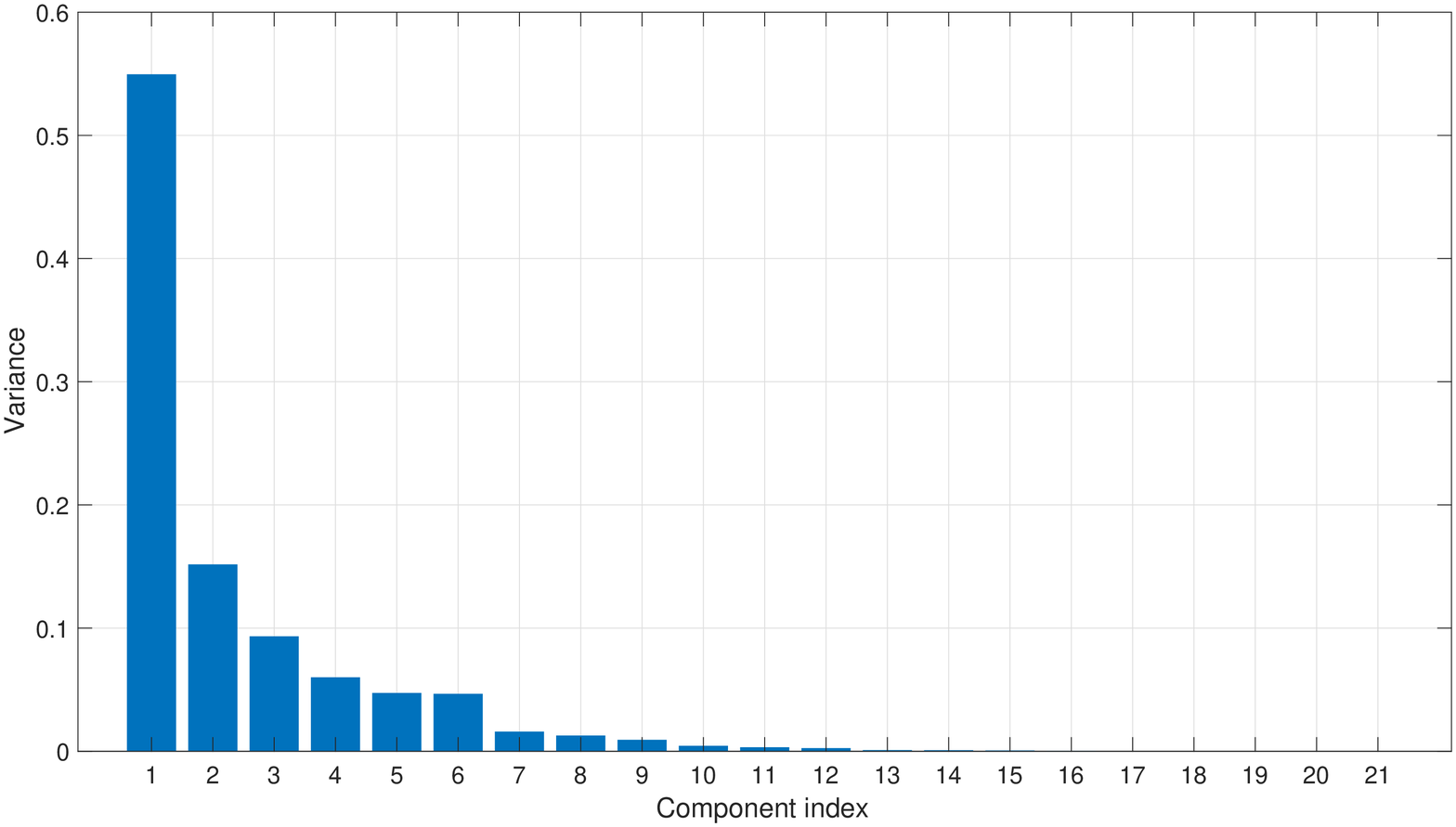}
\label{fig_fe_exam_b}
}
\caption{The variance values of components projected from (a) Dataset A in the third split case, and (b) Dataset B in the second split case.}
\end{figure}

Figure \ref{fig_fe_exam_a} shows the variance values of components after projecting the Dataset A without removing any features in the first split case. It gave three groups of principal components: (part 1), (part 1, part 2) and (part 1, part 2, part 3). Thus we had three choices on the number of retainable principal components: [14, 7, 4]. All of the retainable principal components for each feature-removed dataset are listed in the fourth column of the upper panel of Table \ref{tab_fs_fe}.

With the same philosophy, we display the similar items in Figure \ref{fig_fs_exam_b} and Figure \ref{fig_fe_exam_b} for Dataset B, and the retainable principal components for each feature-removed dataset are listed in the fourth column of the lower panel of Table \ref{tab_fs_fe}.

As for a particular ML model: `RF\_reduced', `RF\_pipe', `MLP' or `GAN', our idea was to span over a wide range of hyper-parameters values and took the model that had the best performance on the test dataset to be the candidate model. Even though it was impossible to test all the hyper-parameters, we supposed that the results in our work were good enough with the optional hyper-parameters listed in Table \ref{tab_ml_parameters}.

\subsection{Identification Results}
The identification results and model hyper-parameters for each split case in our two missions are listed in Table \ref{tab_modelA} and \ref{tab_modelB} respectively, where the Best Models are highlighted by gray color. The bold results are the highest scores under the respective metric.

Since in Mission A we required precise AGN candidates identification, we focused both on precision and accuracy metrics. The candidate models (`RF\_reduced', `RF\_pipe', `MLP', `GAN' and soft ensemble model) were evaluated on five split test datasets. The results showed that the MLP and soft ensemble model had the highest accuracy both in the first and third split cases, and the MLP model in the first split case had the highest precision. Under the consideration that the soft ensemble model has stronger generalization ability and the individual models in the third split case had less parameters, we chose the soft ensemble model of the third split case as Best Model A, where the precision was 97.03\% and accuracy was 97.04\%. While in Mission B, the accuracy was taken as the metric, then the soft ensemble model in the second split case was seen as Best Model B, where the accuracy is 92.13\%.

We also noticed that the results in different split cases were comparable. This means that the Best Models were unbiased and could be employed to identify the unassociated and uncertain sources effectively, as reported in Table \ref{tab_unassociated} and \ref{tab_bcu}, respectively. In order to further understand the model performance, we also recorded the confusion matrices of the Best Model A and B in Table \ref{tab_matrixA} and \ref{tab_matrixB}, respectively. The few misclassified sources in the tables implied the strong capability of Best Models on 3FGL source identification.

\subsection{Physical Analysis}
\label{sucsect_physic_analysis}
\begin{figure}[htbp]
\centering
\subfigure[]{
\includegraphics[width=4in]{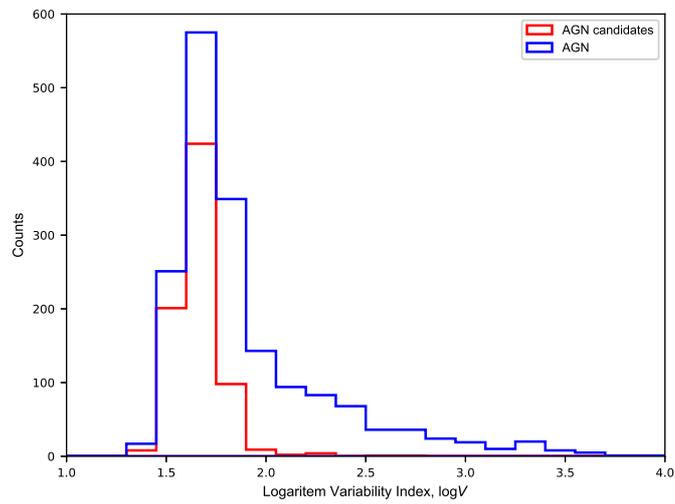}
\label{fig_variability_agn}
}
\subfigure[]{
\includegraphics[width=4in]{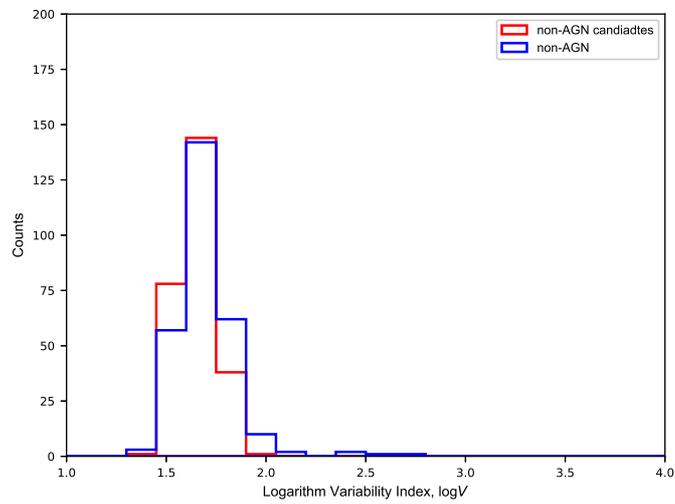}
\label{fig_variability_nonagn}
}
\caption{The (a) AGN and (b) non-AGN logarithm variability index distributions of this work (red) and 3FGL (blue).}
\end{figure}

The variability index is a statistical parameter that indicates the confidence level on the variability of a $\gamma$-ray source \citep{2FGL,3FGL}. Considering that different types of sources should show different variability patterns, in order to cross-check our identification results, we used the variability index as a physical metric to compare the results in Mission A with the associated 3FGL results. The average AGN logarithm variability index ${\rm log} V^{\rm TW}$ of this work is \textbf{$\langle 1.66 \pm 0.12 \rangle$} base on our 748 AGN candidates, while the average logarithm variability index ${\rm log} V^{\rm 3FGL}$ of 1744 3FGL known AGNs is $\langle 1.91 \pm 0.42 \rangle$, as shown in Figure \ref{fig_variability_agn}. The average non-AGNs logarithm variability index ${\rm log} V^{\rm TW}$ of this work is $\langle 1.65 \pm 0.09 \rangle$ base on our 262 non-AGN candidates, while the average logarithm variability index ${\rm log} V^{\rm 3FGL}$ of 280 3FGL known non-AGNs is $\langle 1.68 \pm 0.13 \rangle$, see Figure \ref{fig_variability_nonagn}.

\begin{figure}[htbp]
\centering
\subfigure[]{
\includegraphics[width=4in]{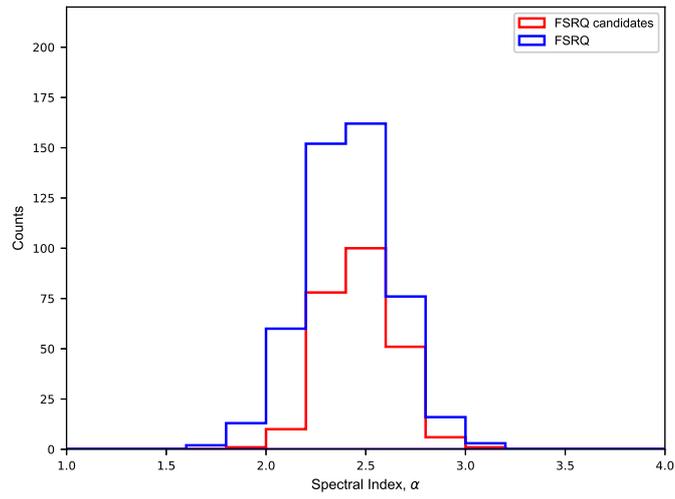}
\label{fig_spectral_fsrq}
}
\subfigure[]{
\includegraphics[width=4in]{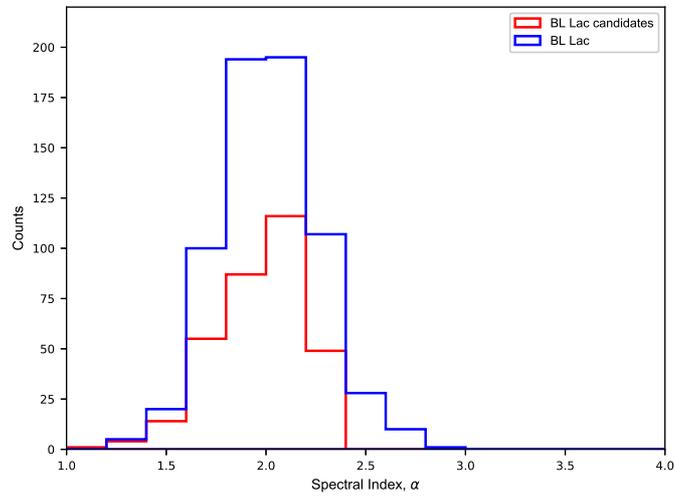}
\label{fig_spectral_bl}
}
\caption{The (a) FSRQ and (b) BL Lac spectral index distributions of this work (red) and 3FGL (blue).}
\end{figure}

Since variability index is not a proper choice to distinguish an FSRQ from BL Lacs, we used the spectral index to cross-check our results. The averaged FSRQ spectral index $\alpha^{\rm TW}$ of this study is $\langle 2.47 \pm 0.17 \rangle$ base on our 247 FSRQ candidates, while the averaged spectral index $\alpha^{\rm 3FGL}$ of 484 3FGL known FSRQs is $\langle 2.42 \pm 0.22 \rangle$, see Figure \ref{fig_spectral_fsrq}. The average BL Lac spectral index $\alpha^{\rm TW}$ of this work was $\langle 1.98 \pm 0.22 \rangle$ base on our 326 BL Lac candidates, while the average spectral index $\alpha^{\rm 3FGL}$ of 660 3FGL known BL Lacs is $\langle 2.02 \pm 0.25 \rangle$, see Figure \ref{fig_spectral_bl}.

\section{Discussion}
\label{sect_discussion}
\subsection{Discussion of our results}
The main goal of this study is to propose a powerful and efficient method to identify the unassociated and uncertain sources in 3FGL. Two dimensionality reduction methods were used to decrease the computational complexity, and the ensemble method was applied to the independently training models for better performance. As for a number of hyper-parameters in each step, grid search method was utilized to decide the final best predictor.

The feature importance distributions of two missions were shown in Figure \ref{fig_fs_exam}, which indicated different identification criterions. In Mission A, the most important feature is F20 (Variability Index). While in Mission B, it is F8 (Power-law Index). This is obviously in line with the AGN's properties and exactly what we expected, for the reason that one of the distinctive properties of AGN is broadband variability, and the different radiation mechanisms of FSRQs and BL Lacs could explain their spectrum differences which behave as power-law index differences.

\begin{figure}[htbp]
\centering
\subfigure[]{
\includegraphics[width=4in]{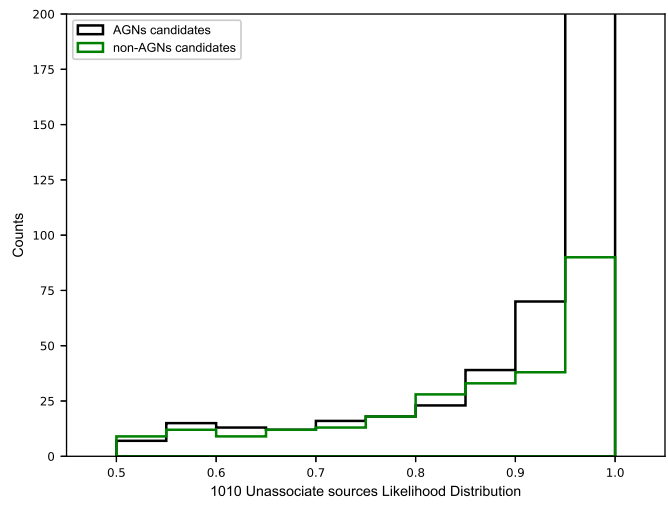}
\label{fig_1010_dis}
}
\subfigure[]{
\includegraphics[width=4in]{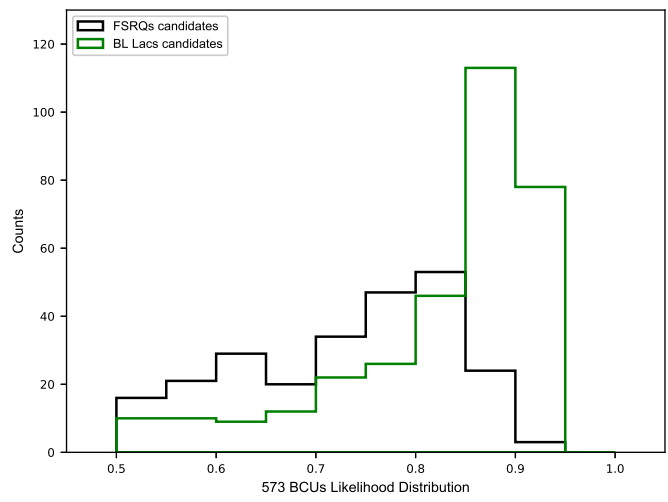}
\label{fig_573_dis}
}
\caption{The probability distributions in (a) Mission A and (b) Mission B.}
\end{figure}

The likelihood distributions of identified AGN candidates and non-AGN candidates in Mission A are shown in Figure \ref{fig_1010_dis}. In this figure, there are 22 AGN candidates and 21 non-AGN candidates with likelihood value lower than 0.6. And, there are 12 AGN candidates and 16 non-AGN candidates with latitude $|b|<10^{\circ}$. It can be attributed that the $\gamma$-ray emissions from the Galactic plane, especially from PSRs and supernova remnants (SNRs), are significant.

For Mission B, the likelihood distribution of identified FSRQ candidates and BL Lac candidates are shown in Figure \ref{fig_573_dis}. From the figure, there are 37 FSRQ candidates and 20 BL Lac candidates whose likelihood are lower than 0.6. In addition, there are 7 FSRQ candidates and 3 BL Lac candidates with latitude $|b|<10^{\circ}$. The best predictor for FSRQs and BL Lacs is the optical spectroscopy observation of emission lines. But for BCUs the optical spectrum is not available or not sensitive enough, then a ML method with outstanding performance has proven to be promising. Indeed, we will plan an optical observation for those FSRQ candidates and BL Lac candidates with likelihood less than 0.7 in our Mission B.

A comparison of variability index and the spectral index and 3FGL is shown in Table \ref{tab_com3f}. We found that our results were consistent with 3FGL except for the average variability index for AGNs. Referring to Figure \ref{fig_variability_agn}, the difference of average variability index is caused by the `long tail' in the distribution of 3FGL AGNs. And this `long tail' can be attributed by some variant AGNs, especially Blazars. These qualitative comparison showing that our results was not wrong, at least without significant mistakes, and implying that our method is functional and efficient to this identification.

\subsection{Comparison with other studies}
There are 708 3FGL unassociated sources can be found in 4FGL. Out of them, 537 sources were identified as AGNs and 171 sources were identified as non-AGNs in our Mission A. In the 537 predicted AGNs, 256 sources were confirmed as AGNs (1 AGN, 191 BCUs, 42 BL Lacs, 19 FSRQs, 3 radio galaxies (RDGs)) by 4FGL. While 18 sources were other certain types (14 pulsars (PSRs), 2 SNRs, 1 Supernova remnant / Pulsar wind nebula (SPP), 1 High-mass binary (HMB)) and 263 sources were still unknown in 4FGL. In the 171 predicted non-AGNs, there were 9 AGNs (8 BCUs and 1 BL Lac), 26 other certain types (1 Magellanic cloud (MC), 18 PSRs, 1 pulsar wind nebula (PWN), 1 SNR, 5 SPP) and 136 unknown in 4FGL.

Moreover, there are 544 3FGL BCUs can be found in 4FGL. Out of them, 221 sources and 323 sources were identified as FSRQs and BL Lacs in our Mission B, respectively. In the 221 predicted FSRQs, 19 sources were confirmed as FSRQs by 4FGL. While 25 sources were other certain types (2 AGN, 21 BL Lacs, 1 HMB, 1 PSR), 11 sources were unknown and 166 sources stayed BCUs in 4FGL. In the 323 predicted BL Lacs, 149 sources were confirmed as BL Lacs by 4FGL. While 9 sources were other certain types ((5 FSRQs, 4 RDGs), 13 sources were unknown and 152 sources stayed BCUs in 4FGL.

The detailed comparison results were enumerated in Table \ref{tab_com4f}. Through the comparison between our results and 4FGL catalogue, we have an accuracy of 91.3\% for Mission A and an accuracy of 83.2\% for Mission B.
Besides, we noticed that our methods were outstanding to identify AGNs in Mission A and identify BL Lacs in Mission B, with precision 93.4\% and 94.3\%, respectively. While the lower precision for non-AGNs and FSRQs identification may result from the quantity of 4FGL confirmed sources with certain types.

In \citet{Parkinson2016}, they collected 3021 3FGL sources, including 1738 AGNs and 166 PSRs. Among the 3021 sources, there were 1008 3FGL sources remain unassociated. These 1008 unassociated sources were identified as 631 AGNs and 377 PSRs with their algorithms. In our Mission A, we identified 748 AGNs and 262 non-AGNs from 1010 3FGL unassociated sources. Through a comparison listed in Table \ref{tab_comp_parkinson}, we found that there were 608 AGN candidates from their work were consistent with ours. But in our work, we focused more on the precision metric. Hence, the identified AGNs with our methods can be considered to be more reliable.

\citet{Chiaro2016} identified 573 BCU sources into 154 FSRQ candidates, 342 BL Lac candidates and 77 stay uncertain. In comparison, we predicted 139 FSRQ candidates and 285 BL Lac candidates in common. 
\citet{Lefaucheur2017} also worked on this identification through ML methods. In their work, 486 BCUs were identified into 146 FSRQ candidates, 295 BL Lac candidates and 45 uncertain sources. Through a comparison, we found 144 FSRQ candidates and 282 BL Lac candidates were in common with theirs, as shown in Table \ref{tab_comp_bcu}.

In addition, both \citet{Chiaro2016} and \citet{Lefaucheur2017} remained several BCUs uncertain which were thoroughly divided into FSRQ and BL Lac in our Mission B. There were 77 sources staying uncertain in \citet{Chiaro2016} and 45 in \citet{Lefaucheur2017}, with 9 common sources included. We summarized these uncertain sources and predicted them to be 74 FSRQs and 39 BL Lacs, as listed in Table \ref{tab_remained_bcu}. With our method, the uncertain BCUs were further identified.

These comparisons indicated that our method was functionally working well as those applied in other work, but we achieved better performance for both missions.

\section{Conclusion}
\label{sect_conclusion}
ML methods has proven to be a promising approach to process astronomical data and they provide classification based on high-dimensionality patterns that human investigation may miss in the first place. In this paper, we used ML methods to complete the source identification in 3FGL for further astrophysical studies. The FS and PCA techniques indeed helped to develop a more efficient algorithm, and ensemble models performed better on unseen samples. Grid search method was demonstrated to be of help to choose the hyper-parameters. With these ML methods, we have successfully identified 748 AGNs out of 1010 unassociated sources, with an accuracy of 97.04\%. Within 573 BCUs, 326 have been identified as BL Lacs and 247 as FSRQs, with an accuracy of 92.13\%.

\section*{Acknowledgements}
This work is partially supported by the National Natural Science Foundation of China (NSFC 11733001, NSFC U15312\\45),  Natural Science Foundation of Guangdong Province (2017\\A030313011), supports for Astrophysics Key Subjects of Guangdong Province and Guangzhou City. Authors would like to acknowledge the scholarship sponsored by Guangzhou University, China and the excellent research facilities provided by Istituto Nazionale di Fisica Nucleare (INFN) and department of Physics and Astronomy of the University of Padova, Padova, Italy. We also thank Riccardo Rando, Alessandro Paccagnella and Denis Bastieri from Padova university for their constructive and valuable suggestions and comments.

\clearpage
\onecolumn

\begin{table}[h]
\footnotesize
\centering
\caption{21 features in 3FGL.\label{tab_features}}
\begin{tabular}{rll}
\specialrule{0em}{2pt}{2pt}\toprule
Index & Feature & Description \\ \midrule
{F}0{} & LII & Galactic longitude \\
{F}1{} & BII & Galactic latitude \\
{F}2{} & ${\rm Significance}_{\rm 100MeV-300GeV}$ & source significance in $\sigma$ units over the 100 MeV to 300 GeV band \\
{F}3{} & $F_{\rm 1-100 GeV}$ & integral photon flux from 1 to 100 GeV in units of ${\rm ph \cdot cm^{-2} \cdot s}$ \\
{F}4{} & Pivot Energy & energy at which error on differential flux is minimal \\
{F}5{} & Spectral index & best fit photon number power-law index \\
{F}6{} & Beta & curvature parameter, for LogParabola \\
{F}7{} & Cutoff Energy & cutoff energy for PL(Super)ExpCutoff \\
{F}8{} & Power law index & best fit power-law index \\
{F}9{} & $F_{\rm 100-300 MeV}$ & integral photon flux from 100 to 300 MeV \\
{F}10{} & ${\rm Significance}_{\rm 100-300 MeV}$ & significance on 100-300 MeV \\
{F}11{} & $F_{\rm 0.3-1 GeV}$ & integral photon flux from 0.3 GeV to 1 GeV \\
{F}12{} & ${\rm Significance}_{\rm 0.3-1 GeV}$ & significance on 0.3-1 GeV \\
{F}13{} & $F_{\rm 1-3 GeV}$ & integral photon flux from 1 GeV to 3 GeV \\
{F}14{} & ${\rm Significance}_{\rm 1-3 GeV}$ & significance on 1-3 GeV \\
{F}15{} & $F_{\rm 3-10 GeV}$ & integral photon flux from 3 GeV to 10 GeV \\
{F}16{} & ${\rm Significance}_{\rm 3-10 GeV}$ & significance on 3-10 GeV \\
{F}17{} & $F_{\rm 10-100 GeV}$ & integral photon flux from 10 GeV to 100 GeV \\
{F}18{} & ${\rm Significance}_{\rm 10-100 GeV}$ & significance on 10-100 GeV \\
{F}19{} & Sign.Curve & fit improvement between power law and LogParabola or PLExpCutoff \\
{F}20{} & Variability Index & \begin{tabular}[c]{@{}l@{}}sum of 2A-log(Likelihood) difference between the flux fitted in each time\\ interval and the average flux over the full catalog interval\end{tabular} \\ \bottomrule
\end{tabular}
\end{table}

\begin{table}[h]\footnotesize
\centering
\caption{RF model used to calculate the FI. Other parameters were set as default in scikit-learn package.\label{tab_fs_parameters}}
\begin{tabular}{ccccccc}
\specialrule{0em}{2pt}{2pt}\toprule
Number of DTs & 
Criterions \\
\midrule
10000	&	entropy \\
30000	&	gini \\
\bottomrule
\end{tabular}
\end{table}

\begin{table}[h]
\centering
\caption{The removable features and retainable principal components in FS and FE step for two datasets, respectively.}
\label{tab_fs_fe}
\resizebox{15cm}{!}{
\begin{tabular}{ccll}
\specialrule{0em}{2pt}{2pt}\toprule
\multicolumn{1}{c}{Dataset A} & \multicolumn{1}{c}{Group} & \multicolumn{1}{c}{Indices of removable features} & \multicolumn{1}{c}{Number of retainable principal components} \\ \midrule
\multirow{4}{*}{Split 1} & 0 & None (keep 21 features) & 14 / 7 / 4 \\
 & 1 & {(}6, 0, 17, 8{)} & 10 / 4 \\
 & 2 & {(}6, 0, 17, 8, 16, 4, 12, 10, 14, 2, 1, 18{)} & 5 / 3 \\
 & 3 & {(}6, 0, 17, 8, 16, 4, 12, 10, 14, 2, 1, 18, 5, 9, 15, 3, 11{)} & 4 \\ \midrule
\multirow{4}{*}{Split 2} & 0 & None (keep 21 features) & 14 / 7 / 4 \\
 & 1 & {(}16, 17, 6, 8, 0{)} & 10 / 5 \\
 & 2 & {(}16, 17, 6, 8, 0, 4, 12, 14, 10, 2, 1, 18{)} & 5 \\
 & 3 & {(}16, 17, 6, 8, 0, 4, 12, 14, 10, 2, 1, 18, 5, 9, 15, 11, 3{)} & 4 \\ \midrule
\multirow{4}{*}{Split 3} & 0 & None (keep 21 features) & 14 / 7 / 4 \\
 & 1 & {(}6, 0, 8, 17, 16{)} & 10 / 5 \\
 & 2 & {(}6, 0, 8, 17, 16, 10, 12, 4, 1, 14, 2, 15, 18{)} & 5 \\
 & 3 & {(}6, 0, 8, 17, 16, 10, 12, 4, 1, 14, 2, 15, 18, 9, 5, 11, 3{)} & 4 \\ \midrule
\multirow{4}{*}{Split 5} & 0 & None (keep 21 features) & 14 / 9 / 4 \\
 & 1 & {(}6, 17, 8, 0{)} & 10 / 5 \\
 & 2 & {(}6, 17, 8, 0, 16, 4, 1, 14, 12, 10, 18, 2, 15{)} & 5 \\
 & 3 & {(}6, 17, 8, 0, 16, 4, 1, 14, 12, 10, 18, 2, 15, 3, 11, 9, 5{)} & 4 \\ \toprule
\multicolumn{1}{c}{Dataset B} & \multicolumn{1}{c}{Group} & \multicolumn{1}{c}{Indices of removable features} & \multicolumn{1}{c}{Number of retainable principal components} \\ \midrule
\multirow{4}{*}{Split 1} & 0 & None (keep 21 features) & 12 / 6 / 4 \\
 & 1 & {(}14, 1, 3, 6, 7{)} & 10 / 5 \\
 & 2 & {(}14, 1, 3, 6, 7, 10, 16, 12, 2, 13, 15, 0{)} & 7 / 4 \\
 & 3 & {(}14, 1, 3, 6, 7, 10, 16, 12, 2, 13, 15, 0, 17, 19, 9, 11{)} & 5 \\ \midrule
\multirow{4}{*}{Split 2} & 0 & None (keep 21 features) & 12 / 9 / 6 \\
 & 1 & {(}7, 6{)} & 11 / 6 / 4 \\
 & 2 & {(}7, 6, 3, 0, 13, 14, 1, 15, 2, 12, 16{)} & 8 / 6 \\
 & 3 & {(}7, 6, 3, 0, 13, 14, 1, 15, 2, 12, 16, 11, 10, 9, 17, 19{)} & 5 \\ \midrule
\multirow{4}{*}{Split 3} & 0 & None (keep 21 features) & 13 / 9 / 6 \\
 & 1 & {(}15, 2, 1, 13, 0, 14, 3, 6, 7{)} & 9 / 6 / 4 \\
 & 2 & {(}15, 2, 1, 13, 0, 14, 3, 6, 7, 10, 16, 12{)} & 7 / 5 \\
 & 3 & {(}15, 2, 1, 13, 0, 14, 3, 6, 7, 10, 16, 12, 11, 9, 17, 19{)} & 5 \\ \midrule
\multirow{4}{*}{Split 4} & 0 & None (keep 21 features) & 12 / 9 / 6 \\
 & 1 & {(}1, 3, 6, 7{)} & 10 / 7 / 5 \\
 & 2 & {(}1, 3, 6, 7, 12, 13, 2, 14, 15, 0{)} & 9 / 6 \\
 & 3 & {(}1, 3, 6, 7, 12, 13, 2, 14, 15, 0, 17, 9, 11, 19, 10, 16{)} & 5 \\ \midrule
\multirow{4}{*}{Split 5} & 0 & None (keep 21 features) & 12 / 10 / 6 \\
 & 1 & {(}14, 13, 3, 6, 7{)} & 11 / 9 / 6 \\
& 3 & {(}1, 3, 6, 7, 12, 13, 2, 14, 15, 0, 17, 9, 11, 19, 10, 16{)} & 5 \\ \midrule
\multirow{4}{*}{Split 5} & 0 & None (keep 21 features) & 12 / 10 / 6 \\
 & 1 & {(}14, 13, 3, 6, 7{)} & 11 / 9 / 6 \\
 & 2 & {(}14, 13, 3, 6, 7, 2, 15, 1, 0, 12, 16{)} & 8 / 4 \\
 & 3 & {(}14, 13, 3, 6, 7, 2, 15, 1, 0, 12, 16, 10, 9, 17, 11, 19{)} & 5 \\ \bottomrule
\end{tabular}}
\end{table}

\begin{table}[htbp]\footnotesize
\centering
\caption{The optional hyper-parameters of RF, MLP and GAN models.}
\label{tab_ml_parameters}
\begin{tabular}{rccc}
\specialrule{0em}{2pt}{2pt}\toprule
Hyper-parameter & \multicolumn{1}{c}{RF} & \multicolumn{1}{c}{MLP} & \multicolumn{1}{c}{GAN} \\ \midrule
Criterion & gini / entropy &  &  \\
Number of DTs & 50 / 100 / 300 / 500 &  &  \\ \midrule
Numer of units &  & \begin{tabular}[c]{@{}l@{}}hidden layer 1: 50 / 30\\ hidden layer 2: 20 / 10\end{tabular} & \begin{tabular}[c]{@{}l@{}}Generator:\\ hidden layer 1: 10 / 20\\ hidden layer 2: 30 / 50\\ Discriminator:\\ hidden layer 1: 50 / 30\\ hidden layer 2: 20 / 10\end{tabular} \\ \midrule
Activation function &  & \multicolumn{2}{l}{relu / elu / sigmoid / tanh / lerelu / softplus} \\
Loss function &  & \multicolumn{2}{l}{l2 / expectation / reg\_expectation / chebyshev / hinge / hinge2 / hinge3 / log / log2 / tan / dcs} \\ \bottomrule
\end{tabular}
\end{table}

\clearpage
{\footnotesize
\begin{longtable}{cllccc}
\caption{Identification results and hyper-parameters of different candidate models in five split cases for Mission A. We combined the precision and accuracy as the metric. The hyper-parameters of soft ensemble models were consistent with the respective individual models. $w1$, $w2$, $w3$, $w4$ represent the weights of `RF\_reduced', `RF\_pipe', `MLP' and `GAN', respectively.}
\label{tab_modelA}\\
\toprule
Dataset A & \multicolumn{1}{c}{Models} & \multicolumn{1}{c}{Hyper-parameters} & Sensitivity & Precision & Accuracy \\ \midrule
\multirow{6}{*}{Split 1} & RF\_reduced & \begin{tabular}[c]{@{}l@{}}removed features: Group 1 \\ criterion: entropy\\ number of trees: 50\end{tabular} & 99.05\% & 96.83\% & 96.38\% \\  \cmidrule{2-6}  
 & RF\_pipe & \begin{tabular}[c]{@{}l@{}}removed features: Group 1 \\ number of retained principal components: 10\\ normalization: L2 Normalizer\\ criterion: gini\\ number of trees: 300\end{tabular} & 99.43\% & 96.48\% & 96.38\% \\  \cmidrule{2-6}  
 & MLP & \begin{tabular}[c]{@{}l@{}}removed features: Group 1 \\ normalization: Standardizer\\ stucture: {[}17 30 20 10 2{]}\\ activation function: tanh\\ loss function: hinge3\end{tabular} & 99.24\% & \textbf{97.38\%} & \textbf{97.04\%} \\  \cmidrule{2-6}  
 & GAN & \begin{tabular}[c]{@{}l@{}}removed features: Group 1 \\ normalization: Standardizer\\ stucture of generator: {[}10 10 20 30 17{]}\\ stucture of discriminator: {[}17 50 20 10 3{]}\\ activation function: leaky relu\\ loss function: expectation\end{tabular} & 99.43\% & 97.02\% & 96.88\% \\  \cmidrule{2-6}    
 & Soft ensemble & \begin{tabular}[c]{@{}l@{}}$w_1=0.05$, $w_2=0.07$, $w_3=0.31$, $w_4=0.57$\end{tabular} & 99.62\% & 97.03\% & \textbf{97.04\%} \\ \midrule
\multirow{6}{*}{Split 2} & RF\_reduced & \begin{tabular}[c]{@{}l@{}}removed features: Group 1 \\ criterion: entropy\\ number of trees: 50\end{tabular} & 99.43\% & 96.48\% & 96.38\% \\  \cmidrule{2-6}  
 & RF\_pipe & \begin{tabular}[c]{@{}l@{}}removed features: Group 1 \\ number of retained principal components: 10\\ normalization: L1 Normalizer\\ criterion: entropy\\ number of trees: 100\end{tabular} & 98.47\% & 96.63\% & 95.72\% \\  \cmidrule{2-6}  
 & MLP & \begin{tabular}[c]{@{}l@{}}removed features: Group 0 \\ normalization: Standardizer\\ stucture: {[}21 30 20 10 2{]}\\ activation function: leaky relu\\ loss function: expectation\end{tabular} & 98.09\% & 96.98\% & 95.72\% \\  \cmidrule{2-6}  
 & GAN & \begin{tabular}[c]{@{}l@{}}removed features: Group 1 \\ normalization: Standardizer\\ stucture of generator: {[}10 10 20 50 16{]}\\ stucture of discriminator: {[}16 50 20 10 3{]}\\ activation function: leaky relu\\ loss function: chebyshev\end{tabular} & 98.09\% & 96.80\% & 95.56\% \\  \cmidrule{2-6}
 & Soft ensemble & \begin{tabular}[c]{@{}l@{}}$w_1=0.34$, $w_2=0.25$, $w_3=0.15$, $w_4=0.26$\end{tabular} & 99.43\% & 96.84\% & 96.71\% \\ \midrule
\multirow{6}{*}{Split 3} & RF\_reduced & \begin{tabular}[c]{@{}l@{}}removed features: Group 1 \\ criterion: gini\\ number of trees: 100\end{tabular} & \textbf{99.81\%} & 96.49\% & 96.71\% \\  \cmidrule{2-6}  
 & RF\_pipe & \begin{tabular}[c]{@{}l@{}}removed features: Group 2 \\ number of retained principal components: 5\\ normalization: Max Normalizer\\ criterion: entropy\\ number of trees: 300\end{tabular} & 98.86\% & 96.28\% & 95.72\% \\  \cmidrule{2-6}  
 & MLP & \begin{tabular}[c]{@{}l@{}}removed features: Group 1 \\ normalization: Standardizer\\ stucture: {[}16 50 10 10 2{]}\\ activation function: relu\\ loss function: reg\_expectation\end{tabular} & 98.47\% & 97.36\% & 96.38\% \\  \cmidrule{2-6}  
 & GAN & \begin{tabular}[c]{@{}l@{}}removed features: Group 1 \\ normalization: Standardizer\\ stucture of generator: {[}10 10 10 30 16{]}\\ stucture of discriminator: {[}16 30 20 10 3{]}\\ activation function: leaky relu\\ loss function: expectation\end{tabular} & 99.62\% & 97.03\% & \textbf{97.04\%} \\  \cmidrule{2-6}   
\rowcolor{mygray} & Soft ensemble & \begin{tabular}[c]{@{}l@{}}$w_1=0.01$, $w_2=0.01$, $w_3=0.01$, $w_4=0.97$\end{tabular} & 99.62\% & 97.03\% & \textbf{97.04\%} \\
\midrule\\

\multirow{6}{*}{Split 4} & RF\_reduced & \begin{tabular}[c]{@{}l@{}}removed features: Group 1 \\ criterion: entropy\\ number of trees: 50\end{tabular} & 99.43\% & 95.95\% & 95.89\% \\  \cmidrule{2-6}  
 & RF\_pipe & \begin{tabular}[c]{@{}l@{}}removed features: Group 2 \\ number of retained principal components: 5\\ normalization: Max Normalizer\\ criterion: gini\\ number of trees: 100\end{tabular} & 99.24\% & 95.94\% & 95.72\% \\  \cmidrule{2-6}  
 & MLP & \begin{tabular}[c]{@{}l@{}}removed features: Group 0 \\ normalization: Standardizer\\ stucture: {[}21 50 10 10 2{]}\\ activation function: leaky relu\\ loss function: log\end{tabular} & 99.05\% & 96.83\% & 96.38\% \\  \cmidrule{2-6}  
 & GAN & \begin{tabular}[c]{@{}l@{}}removed features: Group 0 \\ normalization: Standardizer\\ stucture of generator: {[}10 10 20 50 21{]}\\ stucture of discriminator: {[}21 50 10 10 3{]}\\ activation function: elu\\ loss function: l2\end{tabular} & \textbf{99.81\%} & 95.61\% & 95.89\% \\  \cmidrule{2-6}   
 & Soft ensemble & \begin{tabular}[c]{@{}l@{}}$w_1=0.58$, $w_2=0.26$, $w_3=0.09$, $w_4=0.07$\end{tabular} & \textbf{99.81\%} & 96.49\% & 96.71\% \\ \midrule
\multirow{6}{*}{Split 5} & RF\_reduced & \begin{tabular}[c]{@{}l@{}}removed features: Group 1 \\ criterion: gini\\ number of trees: 100\end{tabular} & 99.62\% & 96.13\% & 96.22\% \\  \cmidrule{2-6}  
 & RF\_pipe & \begin{tabular}[c]{@{}l@{}}removed features: Group 1 \\ number of retained principal components: 5\\ normalization: MaxAbsScaler\\ criterion: entropy\\ number of trees: 300\end{tabular} & 99.43\% & 95.42\% & 95.40\% \\  \cmidrule{2-6}  
 & MLP & \begin{tabular}[c]{@{}l@{}}removed features: Group 0 \\ normalization: Standardizer\\ stucture: {[}21 30 10 10 2{]}\\ activation function: elu\\ loss function: hinge3\end{tabular} & 99.43\% & 95.95\% & 95.89\% \\  \cmidrule{2-6}  
 & GAN & \begin{tabular}[c]{@{}l@{}}removed features: Group 0 \\ normalization: Standardizer\\ stucture of generator: {[}10 10 10 50 21{]}\\ stucture of discriminator: {[}21 50 20 10 3{]}\\ activation function: leaky relu\\ loss function: log2\end{tabular} & 98.86\% & 96.64\% & 96.05\% \\  \cmidrule{2-6}  
 & Soft ensemble & \begin{tabular}[c]{@{}l@{}}$w_1=0.01$, $w_2=0.24$, $w_3=0.16$, $w_4=0.59$\end{tabular} & 99.43\% & 96.48\% & 96.38\% \\
\bottomrule
\end{longtable}}

\clearpage
{\footnotesize
\begin{longtable}{cllccc}
\caption{Identification results and hyper-parameters of different candidate models in five split cases for Mission B. We just took the accuracy score as the metric. The hyper-parameters of soft ensemble models were consistent with the respective individual models. $w1$, $w2$, $w3$, $w4$ represent the weights of `RF\_reduced', `RF\_pipe', `MLP' and `GAN', respectively.}
\label{tab_modelB}\\
\toprule
Dataset B & \multicolumn{1}{c}{Models} & \multicolumn{1}{c}{Hyper-parameters} & Sensitivity & Precision & Accuracy \\ \toprule
\multirow{6}{*}{Split 1} & RF\_reduced & \begin{tabular}[c]{@{}l@{}}removed features: Group 3 \\ criterion: entropy\\ number of trees: 50\end{tabular} & 89.90\% & 89.45\% & 88.05\% \\ \cmidrule{2-6} 
 & RF\_pipe & \begin{tabular}[c]{@{}l@{}}removed features: Group 0 \\ number of retained principal components: 6\\ normalization: StandardScaler\\ criterion: gini\\ number of trees: 50\end{tabular} & 91.41\% & 89.16\% & 88.63\% \\ \cmidrule{2-6} 
 & MLP & \begin{tabular}[c]{@{}l@{}}removed features: Group 2 \\ normalization: Standardizer\\ stucture: {[}9 50 10 10 2{]}\\ activation function: leaky relu\\ loss function: chebyshev\end{tabular} & 91.92\% & 90.55\% & 89.80\% \\ \cmidrule{2-6} 
 & GAN & \begin{tabular}[c]{@{}l@{}}removed features: Group 1 \\ normalization: Standardizer\\ stucture of generator: {[}10 10 20 30 16{]}\\ stucture of discriminator: {[}16 30 10 10 3{]}\\ activation function: leaky relu\\ loss function: expectation\end{tabular} & 87.88\% & \textbf{94.05\%} & 89.80\% \\ \cmidrule{2-6}  
 & Soft ensemble & \begin{tabular}[c]{@{}l@{}}$w_1=0.01$, $w_2=0.29$, $w_3=0.52$, $w_4=0.18$\end{tabular} & 91.41\% & 92.35\% & 90.67\% \\ \midrule
\multirow{6}{*}{Split 2} & RF\_reduced & \begin{tabular}[c]{@{}l@{}}removed features: Group 0 \\ criterion: entropy\\ number of trees: 500\end{tabular} & 91.41\% & 91.88\% & 90.38\% \\\cmidrule{2-6} 
 & RF\_pipe & \begin{tabular}[c]{@{}l@{}}removed features: Group 1 \\ number of retained principal components: 11\\ normalization: Standardizer\\ criterion: gini\\ number of trees: 300\end{tabular} & 90.40\% & 92.27\% & 90.09\% \\ \cmidrule{2-6} 
 & MLP & \begin{tabular}[c]{@{}l@{}}removed features: Group 3 \\ normalization: Standardizer\\ stucture: {[}5 50 10 10 2{]}\\ activation function: tanh\\ loss function: chebyshev\end{tabular} & 92.42\% & 93.37\% & 91.84\% \\ \cmidrule{2-6} 
 & GAN & \begin{tabular}[c]{@{}l@{}}removed features: Group 3 \\ normalization: Standardizer\\ stucture of generator: {[}10 10 20 50 5{]}\\ stucture of discriminator: {[}5 30 10 10 3{]}\\ activation function: elu\\ loss function: l2\end{tabular} & 92.42\% & 92.89\% & 91.55\% \\ \cmidrule{2-6} 
\rowcolor{mygray} & Soft ensemble & \begin{tabular}[c]{@{}l@{}}$w_1=0.01$, $w_2=0.19$, $w_3=0.73$, $w_4=0.07$\end{tabular} & 92.42\% & 93.85\% & \textbf{92.13\%} \\ \midrule
\multirow{6}{*}{Split 3} & RF\_reduced & \begin{tabular}[c]{@{}l@{}}removed features: Group 0 \\ criterion: entropy\\ number of trees: 100\end{tabular} & 91.41\% & 90.05\% & 89.21\% \\  \cmidrule{2-6} 
 & RF\_pipe & \begin{tabular}[c]{@{}l@{}}removed features: Group 1 \\ number of retained principal components: 6\\ normalization: Max Normalizer\\ criterion: gini\\ number of trees: 100\end{tabular} & 92.93\% & 88.89\% & 89.21\% \\ \cmidrule{2-6} 
 & MLP & \begin{tabular}[c]{@{}l@{}}removed features: Group 2 \\ normalization: Standardizer\\ stucture: {[}9 30 10 10 2{]}\\ activation function: sigmoid\\ loss function: hinge2\end{tabular} & 91.41\% & 91.41\% & 90.09\% \\ \cmidrule{2-6} 
 & GAN & \begin{tabular}[c]{@{}l@{}}removed features: Group 2 \\ normalization: Standardizer\\ stucture of generator: {[}10 10 10 50 9{]}\\ stucture of discriminator: {[}9 50 10 10 3{]}\\ activation function: elu\\ loss function: log2\end{tabular} & 91.41\% & 91.88\% & 90.38\% \\ \cmidrule{2-6}
 & Soft ensemble & \begin{tabular}[c]{@{}l@{}}$w_1=0.01$, $w_2=0.43$, $w_3=0.02$, $w_4=0.54$\end{tabular} & \textbf{93.43\%} & 91.13\% & 90.96\%\\
\midrule\\

\multirow{6}{*}{Split 4} & RF\_reduced & \begin{tabular}[c]{@{}l@{}}removed features: Group 2 \\ criterion: gini\\ number of trees: 100\end{tabular} & 91.41\% & 90.96\% & 89.80\% \\ \cmidrule{2-6} 
 & RF\_pipe & \begin{tabular}[c]{@{}l@{}}removed features: Group 1 \\ number of retained principal components: 10\\ normalization: L1 Normalizer\\ criterion: entropy\\ number of trees: 300\end{tabular} & 87.88\% & 92.55\% & 88.92\% \\ \cmidrule{2-6} 
 & MLP & \begin{tabular}[c]{@{}l@{}}removed features: Group 3 \\ normalization: Standardizer\\ stucture: {[}5 30 20 10 2{]}\\ activation function: tanh\\ loss function: hinge2\end{tabular} & 90.91\% & 91.84\% & 90.09\% \\ \cmidrule{2-6} 
 & GAN & \begin{tabular}[c]{@{}l@{}}removed features: Group 1 \\ normalization: Standardizer\\ stucture of generator: {[}10 10 10 50 17{]}\\ stucture of discriminator: {[}17 50 20 10 3{]}\\ activation function: relu\\ loss function: tan\end{tabular} & 91.41\% & 92.35\% & 90.67\% \\ \cmidrule{2-6} 
 & Soft ensemble & \begin{tabular}[c]{@{}l@{}}$w_1=0.02$, $w_2=0.01$, $w_3=0.17$, $w_4=0.80$\end{tabular} & 91.92\% & 92.86\% & 91.25\% \\ \midrule
\multirow{6}{*}{Split 5} & RF\_reduced & \begin{tabular}[c]{@{}l@{}}removed features: Group 1 \\ criterion: gini\\ number of trees: 50\end{tabular} & 92.42\% & 88.41\% & 88.63\% \\ \cmidrule{2-6} 
 & RF\_pipe & \begin{tabular}[c]{@{}l@{}}removed features: Group 2 \\ number of retained principal components: 4\\ normalization: Standardizer\\ criterion: gini\\ number of trees: 100\end{tabular} & 92.42\% & 87.56\% & 88.05\% \\ \cmidrule{2-6} 
 & MLP & \begin{tabular}[c]{@{}l@{}}removed features: Group 0 \\ normalization: Standardizer\\ stucture: {[}21 50 10 10{]}\\ activation function: relu\\ loss function: reg\_expectation\end{tabular} & 89.39\% & 91.24\% & 88.92\% \\ \cmidrule{2-6} 
 & GAN & \begin{tabular}[c]{@{}l@{}}removed features: Group 0 \\ normalization: Standardizer\\ stucture of generator: {[}10 10 10 30 21{]}\\ stucture of discriminator: {[}21 30 20 10 3{]}\\ activation function: leaky relu\\ loss function: hinge3\end{tabular} & 87.88\% & 91.58\% & 88.34\% \\ \cmidrule{2-6} 
 & Soft ensemble & \begin{tabular}[c]{@{}l@{}}$w_1=0.56$, $w_2=0.38$, $w_3=0.05$, $w_4=0.01$\end{tabular} & 91.41\% & 90.50\% & 89.50\% \\
\bottomrule
\end{longtable}}

\begin{table}[h]\footnotesize
\centering
\caption{Identification results of the 1010 unassociated sources with Best Model A.}
\label{tab_unassociated}
\begin{threeparttable}
\begin{tabular}{rccccccc}
\specialrule{0em}{2pt}{2pt}
\toprule
 3FGL name      &  $l(^{\circ} )$  & $b(^{\circ} )$ & Spectral index & Variability index & Prediction  & Likelihood  \\
\midrule
J0000.1+6545	&	117.69	&	3.4		&	2.411	&	40.754	&	non-AGN		&	0.56142	\\
J0000.2-3738	&	345.41	&	-74.95	&	1.867	&	44.845	&	AGN		&	0.98165	\\
J0001.6+3535	&	111.66	&	-26.19	&	2.351	&	32.394	&	AGN		&	0.95838	\\
J0002.0-6722	&	310.13	&	-49.07	&	1.946	&	46.125	&	AGN		&	0.97277	\\
J0002.6+6218	&	117.3	&	-0.04		&	2.086	&	48.024	&	AGN		&	0.81137	\\
...			& 	... 		& 	...		&	...		&	...		&	...		&	...		\\
\bottomrule
\end{tabular}
		\begin{tablenotes}
			\item[*] The full version of this table is available online.
		\end{tablenotes}
	\end{threeparttable}
\end{table}

\begin{table}[h]\footnotesize
\centering
\caption{Identification results of the 573 BCUs with Best Model B.}
\label{tab_bcu}
\begin{threeparttable}
\begin{tabular}{rccccccc}
\specialrule{0em}{2pt}{2pt}
\toprule
 3FGL name      &  $l(^{\circ} )$  & $b(^{\circ} )$ & Spectral index & Variability index & Prediction  & Likelihood  \\
\midrule
J0002.2-4152	&	334.07	&	-72.15	&	2.089	&	56.351	&	BL Lac	&	0.80919	\\
J0003.2-5246	&	318.97	&	-62.83	&	1.895	&	45.28	&	BL Lac	&	0.90905	\\
J0003.8-1151	&	84.43	&	-71.09	&	2.024	&	48.492	&	BL Lac	&	0.84678	\\
J0009.6-3211	&	1.26		&	-79.61	&	2.31		&	42.534	&	BL Lac	&	0.52062	\\
J0012.4+7040	&	119.68	&	8.04		&	2.484	&	51.576	&	FSRQ	&	0.73085	\\
...			& 	... 		& 	...		&	...		&	...		&	...	&	...		\\
\bottomrule
\end{tabular}
		\begin{tablenotes}
			\item[*] The full version of this table is available online.
		\end{tablenotes}
	\end{threeparttable}
\end{table}

\begin{table}[h]\footnotesize
\centering
\caption{Confusion matrix of Best Model A on the third split test dataset.}
\label{tab_matrixA}
\begin{tabular}{cccc}
\specialrule{0em}{2pt}{2pt}\toprule
\multicolumn{2}{c}{\multirow{2}{*}{Number of Predicted Results}} & \multicolumn{2}{c}{Predicted} \\ \cmidrule{3-4} 
\multicolumn{2}{c}{} & AGN & non-AGN \\ \midrule
\multirow{2}{*}{Actual} & AGN & 522 (TP) & 2 (FN) \\
 & non-AGN & 16 (FP) & 68 (TN) \\ \bottomrule
\end{tabular}
\end{table}

\begin{table}[h]\footnotesize
\centering
\caption{Confusion matrix of Best Model B on the second split test dataset.}
\label{tab_matrixB}
\begin{tabular}{cccc}
\specialrule{0em}{2pt}{2pt}\toprule
\multicolumn{2}{c}{\multirow{2}{*}{Number of Predicted Results}} & \multicolumn{2}{c}{Predicted} \\ \cmidrule{3-4} 
\multicolumn{2}{c}{} & BL Lac & FSRQ \\ \midrule
\multirow{2}{*}{Actual} & BL Lac & 183 (TP) & 15 (FN) \\
 & FSRQ & 12 (FP) & 133 (TN) \\ \bottomrule
\end{tabular}
\end{table}

\begin{table}[htbp]\footnotesize
\centering
\caption{Comparison of parameters from this work and 3FGL.}
\label{tab_com3f}
\begin{tabular}{ccccc}
\specialrule{0em}{2pt}{2pt}\toprule
Mission & Parameter & Class &  This Work  & 3FGL \\ \midrule
\multirow{2}{*} {A} & Variability & AGN & $1.66 \pm 0.12$ & $1.91 \pm 0.42$  \\   \cmidrule{3-5}
   & index & non-AGN & $1.65 \pm 0.09$ & $1.68 \pm 0.13$  \\   \midrule
\multirow{2}{*} {B} & Spectral & FSRQ & $2.47 \pm 0.17$ & $2.42 \pm 0.22$  \\   \cmidrule{3-5}
   & index & BL Lac & $1.98 \pm 0.22$ & $2.02 \pm 0.25$  \\   \bottomrule
\end{tabular}
\end{table}

\begin{table}[htbp]\footnotesize
\renewcommand\arraystretch{1.2}
\centering
\caption{Comparison of our results with 4FGL.}
\label{tab_com4f}
\begin{threeparttable}
\begin{tabular}{c|ccc|c|c}
\specialrule{0em}{2pt}{2pt} \toprule 
Prediction of Mission A &  & 4FGL Confirmation &   & Precision & Accuracy \\ \midrule
537 AGNs & 256 AGNs & 18 Others & 263 Un & $\frac{256}{256+18} = 93.4\%$ & \multirow{2}{*} {$\frac{256+26}{256+26+18+9} = 91.3\%$}\\
171 non-AGNs & 9 AGNs & 26 Others &  136 Un  & $\frac{26}{26+9} = 74.3\%$ &  \\  \midrule
Prediction of Mission B &  & 4FGL Confirmation &   & Precision & Accuracy \\ \midrule

221 FSRQs & 19 FSRQs & 25 Others & 11 Un + 166 BCUs & $\frac{19}{19+25} = 43.2\%$ &  \multirow{2}{*} {$\frac{19+149}{19+149+25+9} = 83.2\%$}   \\
323 BL Lacs & 149 BL Lacs & 9 Others & 13 Un + 152 BCUs & $\frac{149}{149+9} = 94.3\%$ & \\ \bottomrule
\end{tabular}
		\begin{tablenotes}
			\item[1] `Un' means unknown type of sources.
		\end{tablenotes}
	\end{threeparttable}
\end{table}

\begin{table}[h]\footnotesize
\centering
\caption{Comparisons of our results with \citet{Parkinson2016} in Mission A.}
\label{tab_comp_parkinson}
\begin{tabular}{rccc}
\specialrule{0em}{2pt}{2pt}\toprule
Class & Our result & P2016 & Common  \\ 
\midrule
AGN & 748 & 631 & 608  \\ 
non-AGN & 262 & ~ & ~ \\ 
PSR & ~ & 377 &  ~ \\  \midrule
In total & 1010 & 1008 & ~ \\ \bottomrule
\end{tabular}
\end{table}

\begin{table}[h]\footnotesize
\centering
\caption{Comparisons of our results with \citet{Chiaro2016} and \citet{Lefaucheur2017} in Mission B.}
\label{tab_comp_bcu}
\begin{tabular}{rccccc}
\specialrule{0em}{2pt}{2pt}\toprule
Class & Our result & C2016 & Common & L2017 &  Common \\ 
\midrule
FSRQ & 247 & 154 & 139 & 146 & 144   \\ 
BL Lac & 326 & 342 & 285 & 295 & 282    \\
Uncertain & 0 & 77 & ~ & 45 & ~   \\  \midrule
In total & 573 & 573 & ~ & 486 & ~   \\ \bottomrule
\end{tabular}
\end{table}

\begin{table}[h]\footnotesize
\centering
\caption{113 uncertain BCUs from \citet{Chiaro2016} and \citet{Lefaucheur2017}.}
\label{tab_remained_bcu}
\begin{threeparttable}
\begin{tabular}{cccccc}
\toprule
3FGL name       & C2016	&  L2017   &    TW     & Likelihood    \\
\midrule
J0002.2-4152	&		&	Unc	&	BL Lac	&	0.80919	\\
J0021.6-6835	&	Unc	&		&	FSRQ	&	0.8107	\\
J0050.0-4458	&	Unc	&		&	FSRQ	&	0.78408	\\
J0055.2-1213	&	Unc	&		&	FSRQ	&	0.69895	\\
J0059.1-5701	&	Unc	&		&	FSRQ	&	0.82588	\\
...			& 	... 	& 	...	&	...	&	...		\\
\bottomrule
\end{tabular}
		\begin{tablenotes}
			\item[*] The full version of this table is available online;
			\item[1] `Unc' means `Uncertain type of BCUs';
			\item[2] The last two columns indicate the source prediction and likelihood in our work.
		\end{tablenotes}
	\end{threeparttable}
\end{table}

\end{document}